\documentclass[11pt,a4paper]{article}
\usepackage{jheppub}
\usepackage{epsfig,bbm,cancel}
\usepackage{slashed,xcolor}
\usepackage[normalem]{ulem} 
\usepackage{amssymb}

\newcommand{\tr}{\text{Tr}}

\title{Are There BPS Dyons in The Generalized $SU(2)$ Yang-Mills-Higgs Model?}
\author{Ardian Nata Atmaja}
\affiliation{Research Center for Quantum Physics, National Research and Innovation Agency (BRIN), Kompleks PUSPIPTEK Serpong, Tangerang 15310, Indonesia.}
\emailAdd{ardi002@brin.go.id}

\abstract{
We use the well-known Bogomolny's equations, in general coordinate system, for BPS monopoles and dyons in the $SU(2)$ Yang-Mills-Higgs model to obtain an explicit form of BPS Lagrangian density under the BPS Lagrangian method. We then generalize this BPS Lagrangian density and use it to derive several possible generalized Bogomolny's equations, with(out) additional constraint equations, for BPS monopoles and dyons in the generalized $SU(2)$ Yang-Mills-Higgs model. We also compute the stress-energy-momentum tensor of the generalized model, and argue that the BPS monopole and dyon solutions are stable if all components of the stress-tensor density are zero in the BPS limit. This stability requirement implies the scalar fields-dependent couplings to be related to each other by an equation, which is different from the one obtained in~\cite{Atmaja:2018cod}, and then picks particular generalized Bogomolny's equations, with no additional constraint equation, out of those possible equations. We show that the computations in~\cite{Atmaja:2018cod} are actually incomplete. Under the Julia-Zee ansatz, the generalized Bogomolny's equations imply all scalar fields-dependent couplings must be constants, whose solutions are the BPS dyons of the $SU(2)$ Yang-Mills-Higgs model~\cite{Prasad:1975kr}, or in another words there are no generalized BPS dyon solutions under the Julia-Zee ansatz. We propose two possible ways for obtaining generalized BPS dyons, where at least one of the scalar fields-dependent couplings is not constant, that are by using different ansatze, such as axially symmetric ansatz for higher topological charge dyons; and/or by considering the most general BPS Lagrangian density.}


\begin{document}

\maketitle
\thispagestyle{empty}
\setcounter{page}{1}

\section{Introduction}

The natural extension of monopoles are dyons which are essentially monopoles with non-zero electrical charges. They were first proposed as an alternative to quarks by Julian Schwinger~\cite{Schwinger:1969ib}, whose quantum mechanical properties were studied by Zwanziger~\cite{Zwanziger:1969by,Zwanziger:1968rs}. Like monopoles, it is also natural for dyons to exist in the non-Abelian gauge theories. The first example of monopoles existence was shown in the $SU(2)$ Yang-Mills-Higgs model, also known as Georgi-Glashow model~\cite{Georgi:1972cj}, by Polyakov and 't Hooft~\cite{Polyakov:1974ek,tHooft:1974kcl}. It was later shown that the dyons could also exist in the same model by Julia and Zee~\cite{Julia:1975ff}. 

The explicit solutions of 'tHooft-Polyakov monopoles and Julia-Zee dyons were presented by Prasad and Sommerfield by taking a special limit to the model~\cite{Prasad:1975kr}. These solutions turn out to be solutions of first-order differential equations, known as Bogomolny's equations, that were derived by Bogomolny~\cite{Bogomolny:1975de}\footnote{These solutions to Bogomolny's equations are generally called BPS solutions for monopoles and dyons, or briefly called BPS monopoles and BPS dyons.}. The solutions saturate the non-trivial static energy bound which turns out to be proportional with the topological charge. Obtaining the Bogomolny's equations of a model is important in particular to study the topological stability of its solitons solutions. There have been some methods developed in these directions which are the first-order formalism~\cite{Bazeia:2005tj,Bazeia:2007df}, FOEL (First-Order-Euler-Lagrange) formalism by using the concept of strong necessary conditions~\cite{Sokalski:2001wk,Adam:2016ipc}, the {\it On-Shell} method~\cite{Atmaja:2014fha,Atmaja:2015lia}, and the BPS Lagrangian method~\cite{Atmaja:2015umo,Atmaja:2018ddi}.

The new studies on monopoles and dyons have been carried out recently that give arise to new features and dynamics. Some of those studies were based on modifications to the $SU(2)$ Yang-Mills-Higgs model~\cite{Georgi:1972cj}. One of the studies were done by inserting extra degrees of freedom along with additional global symmetries~\cite{Shifman:2015ama}. The other one is by adding scalar fields-dependent couplings to each of its kinetic terms, that we shall call as generalized $SU(2)$ Yang-Mills-Higgs model proposed in~\cite{Casana:2012un}, in which the monopoles could be endowed with internal structures~\cite{Bazeia:2018fhg}. In condensed matter, this effective model could be important for the magnetic materials known as spin ice that has a capability to support exotic magnetic structures such as monopoles~\cite{PhysRevLett.79.2554,1999Natur.399..333R,Bramwell1495}. There is also a possibility of electric dipole existing in these monopoles~\cite{Khomskii2012}. This motivates us to study further about dyons in this effective model which may exist as exotic structures in spin ice.  More recently, there is also a study by combining these two modifications that leads to monopoles endowed with some internal structures~\cite{Bazeia:2018eta}. 

In this article we would like to study about BPS dyons in the generalized $SU(2)$ Yang-Mills model. In the generalized $SU(2)$ Yang-Mills-Higgs model the dynamics of overall system may differ from its corresponding canonical model, which are explicitly shown in the generalized Bogomolny's equations for BPS monopoles and for dyons. The first-order formalism has been used to derive the generalized Bogomolny's equations for BPS monopoles in which the solutions are called generalized BPS monopoles~\cite{Casana:2012un}. These Bogomolny's equations exist only if the scalar fields-dependent couplings are related to each other by an equation. On the other hand, the BPS Lagrangian method has been used to rederive the Bogomolny's equations for BPS monopoles and it also managed to obtain the Bogomolny's equations for BPS dyons, which exist only if the scalar fields-dependent couplings are related to each other by a more general equation~\cite{Atmaja:2018cod}. However, all those derivations rely on a particular hedgehog ansatz namely 't Hooft-Polyakov and Julia-Zee ansatze for monopoles and dyons respectively. It is then necessary to find the generalized Bogomolny's equations for BPS monopoles and dyons in general coordinate system that are independent of any ansatz in order to study other possible soliton solutions and configurations. We also would like to verify if the relations between scalar fields-dependent couplings for BPS monopoles and dyons, derived in~\cite{Atmaja:2018cod} for the Julia-Zee ansatz, are still hold in the general coordinate system for any ansatz. Nevertheless the existence of these Bogomolny's equations in general coordinate system are inevitable for extending the corresponding model to its supersymmetric version.

For this matter, we will use the BPS Lagrangian method and generalize its procedures in order to work in general coordinate system. At first, we will employ it to the case of the $SU(2)$ Yang-Mills-Higgs model and derive its corresponding BPS Lagrangian density using the fact that we already had the well-known Bogomolny's equations, in the general coordinate system, for BPS monopoles and dyons at our disposal. We generalize the BPS Lagrangian density, by multiplying each term in the BPS Lagrangian density with arbitrary function of the scalar-fields, and then use this generalized BPS Lagrangian density to derive the generalized Bogomolny's equations for BPS monopoles and dyons in the generalised $SU(2)$ Yang-Mills-Higgs model. We write down all possible generalized Bogomolny's equations, that correspond to the generalized BPS Lagrangian density, for BPS monopoles and dyons in the general coordinate system and study their stabilities from the stress-energy-momentum density tensor. As an example we will apply the Julia-Zee ansatz into the generalized Bogomolny's equations, along with the constraint equations, and compare the results with the ones in~\cite{Atmaja:2018cod}.

\section{The Generalized $SU(2)$ Yang-Mills-Higgs Model}

In this article we will consider the generalized $SU(2)$ Yang-Mills-Higgs model with the following Langrangian density~\cite{Casana:2012un,Atmaja:2018cod}\footnote{Here we follow the notations in~\cite{Atmaja:2018cod}.}: 
 \begin{equation}\label{gen YMH}
  \mathcal{L}=-{w(|\Phi|)\over 2}\tr\left(F_{\mu\nu}F^{\mu\nu}\right)+ G(|\Phi|)\tr\left(D_\mu\Phi D^\mu\Phi\right)-V(|\Phi|),
 \end{equation}
where $w,G>0$ and $V\geq 0$ are functions of scalar fields and are also $SU(2)$ invariant, with $|\Phi|=2\tr\left(\Phi\right)^2$, $F_{\mu\nu}=\partial_\mu A_\nu-\partial_\nu A_\mu-ie \left[A_\mu,A_\nu\right]$, $D_\mu\equiv\partial_\mu-ie\left[A_\mu,\right]$, and $\mu,\nu=0,1,2,3$ are spacetime indices with metric signature $(+---)$. In terms of components, the gauge and scalar fields are
\begin{equation}
 A_\mu={1\over 2}\tau^a A^a_\mu,\qquad \Phi={1\over 2}\tau^a \Phi^a,
\end{equation}
with $a=1,2,3$ and $\tau^a$ are the Pauli matrices. The full Euler-Lagrange equations are
\begin{subequations}\label{full EoM}
 \begin{eqnarray}
 D_\mu\left(G D^\mu\Phi\right)&=&2{\partial G\over\partial|\Phi|} \tr\left(D_\mu\Phi D^\mu\Phi\right)\Phi-{\partial w\over\partial|\Phi|} \tr\left(F_{\mu\nu} F^{\mu\nu}\right)\Phi-2{\partial V\over\partial|\Phi|}\Phi,\\
 D_\nu\left(w~ F^{\mu\nu}\right)&=&-ieG[\Phi, D^\mu\Phi].\label{full EoM gauge}
\end{eqnarray}
\end{subequations}

In the literature, the solutions for monopoles and dyons were mostly found by taking the following Julia-Zee ansatz
\begin{subequations}\label{eq:ansatz}
\begin{align}
\Phi^a&= f(r) {x^a\over r},\\
A^a_0&={j(r)\over e} {x^a\over r},\\
A^a_i&={1-a(r)\over e} \epsilon^{aij} {x^j\over r^2},
\end{align}
\end{subequations}
where $x^a\equiv(x,y,z)$, as well as $x^{i,j}\equiv(x,y,z)$, denotes the Cartesian coordinates. Here we shall call the function $f(r),j(r),$ and $a(r)$ as effective fields of the scalar Higgs, the scalar potential, and the vector potential fields respectively. Notes that the Levi-Civita symbol $\epsilon^{aij}$ in (\ref{eq:ansatz}) mixes the spatial indices and the group index. The ansatz (\ref{eq:ansatz}) is actually defined for the Julia-Zee dyons while for the 't Hooft-Polyakov monopoles is defined by taking $j=0$, or known as the 't Hooft-Polyakov ansatz. For the later purposes let us define $E_i={1\over2}\tau^a E^a_i \equiv F_{0i} $ and $B_i={1\over2}\tau^a B^a_i\equiv{1\over 2}\epsilon_{ijk}F_{jk}$ which represent electric fields and magnetic fields respectively.

\section{BPS Lagrangian Method in General Coordinate System}

The standard way to obtain Bogomolny's equations of a model is by considering its energy functional. We then rewrite it by completing the square in such a way that it contains ``boundary'' term. Using this Bogomolny's trick, we will able to obtain the Bogomolny's equations by minimizing the energy functional of this form~\cite{Bogomolny:1975de}. However the Bogomolny's trick may not be applicable to all models since there is no rigorous way to do it. Nevertheless even if we obtain the Bogomolny's equations, by using the Bogomolny's trick, we still need to verify if these Bogomolny's equations satisfy the full equations of motion trivially, e.g. in the case of $SU(2)$ Yang-Mills-Higgs model the Bogomolny's equations indeed satisfy the Gauss's law constraint equation trivially~\cite{Manton:2004tk,Weinberg:2012pjx}. 

A more rigourous way to obtain the Bogomolny's equations of a model is by using BPS Lagrangian method~\cite{Atmaja:2015umo,Atmaja:2018ddi}. Rather than considering the energy functional, the BPS Lagrangian method works directly to the action of a model. This method uses the fact that the Lagrangian density of most models contains up to quadratic in first-derivative of the fields. As an example Lagrangian density of a model with $N$-scalar fields, $\phi^i$ where $i=1,\ldots,N$, can be rewritten into the following form
\begin{equation}\label{general-lagrangian}
    \mathcal{L} = \sum_{i = 1}^{N}g^i(\phi^j)\left(\partial_{\mu}{\phi^i} - f_\mu^{i}\left(\phi^j,\partial_{\nu}\phi^{j}\right)\right)^{2} + \mathcal{L}_{BPS} ~ ,
\end{equation}
where in general $g^i(\phi^j)$ is a function of fields $\phi^j$'s, and $f_\mu^i$ is a function of fields $\phi^j$'s and their first-derivative $\partial_\nu\phi^j$'s, with $j=1,\ldots,N$, but not of $\partial_\mu\phi^i$. Here we shall call $\mathcal{L}_{BPS}$ as BPS Lagrangian density which in general is a function of fields $\phi^j$'s and their first-derivative $\partial_\nu\phi^j$'s. The Bogomolny's equations are obtained from \eqref{general-lagrangian} in the limit where $\mathcal{L}-\mathcal{L}_{BPS}=0$, or also known as the BPS limit condition, such that
\begin{equation}
 \partial_{\mu}\phi^i = f_\mu^{i}\left(\phi^j,\partial_{\nu}\phi^{j}\right).
\end{equation}
The BPS Lagrangian density plays an important role in the BPS Lagrangian method. Its Euler-Lagrange equations are called constraint equations,
\begin{equation}
 \partial_\mu\left(\delta\mathcal{L}_{BPS} \over \delta(\partial_\mu\phi^i)\right)={\delta\mathcal{L}_{BPS} \over \delta\phi^i},
\end{equation}
which must be considered, in addition to the Bogomolny's equations, in order to find the solitonic solutions. Depending on the choice of terms in the BPS Lagrangian density, its Euler-Lagrange equations could be all trivial. In this case the BPS Lagrangian density contains only ``boundary'' terms such that there are no additional constraint equations. Several possible ``boundary'' terms, that can be included in the BPS Lagrangian density, have been studied in~\cite{Adam:2016ipc}. For most of the known cases, their BPS Lagrangian densities were found to contain only ``boundary'' terms, under some particular ansatze~\cite{Atmaja:2015umo,Atmaja:2018cod}. There has been a study on the BPS Lagrangian density containing ``non-boundary'' terms, under a particular ansatz, which results in solitonic solutions whose stress tensor are non-zero~\cite{Atmaja:2018ddi}. However, the existance ``non-boundary'' terms does not always imply additional constraint equations. In the BPS limit, these constraint equations could be trivially satisfied and thus can be neglected in finding the solitonic solutions. We will see that this is the case for BPS Lagrangian densities considered in this article.

Let us first consider the $SU(2)$ Yang-Mills-Higgs model by taking $G=w=1$ into the Lagrangian density (\ref{gen YMH}), which can be written in terms of $E_i$ and $B_i$ as~\cite{Georgi:1972cj}
\begin{equation}\label{SU(2) YMH}
 \mathcal{L}=\tr\left(E_i\right)^2-\tr\left(B_i\right)^2+\tr\left(D_0\Phi\right)^2-\tr\left(D_i\Phi\right)^2-V,
\end{equation}
where $i=1,2,3$ is the spatial indices. The next step in the BPS Lagrangian method is to write the BPS Lagrangian density. The BPS Lagrangian density initially consisted of terms that are linear in the first-derivative of fields with additional condition that they are ``boundary'' terms, which its Euler-Lagrange equations are trivial~\cite{Atmaja:2015umo}. It was then extended to contain the terms that are quadratic in the first-derivative of fields~\cite{Atmaja:2018ddi}. Furthermore, it can be generalized to contain terms that are polynomial in the first-derivative of fields, or in general terms that are not necessary ``boundary'' terms as such its Euler-Lagrange equations are non-trivial~\cite{Atmaja:2019gce}. These Euler-Lagrange equations will then be constraint equations that must be considered in finding the solutions. However so far the BPS Lagrangian density has been written under certain ansatzs, such as (\ref{eq:ansatz}), in the spherical coordinate system~\cite{Atmaja:2018cod}. Generalizing to general coordinate system would then implies the BPS Lagrangian density with massive terms and hence making the computation to be more complicated. For particular case of the $SU(2)$ Yang-Mills-Higgs model, we will make use of the well-known Bogomolny's equations for monopoles and dyons~\cite{Bogomolny:1975de,Manton:2004tk,Weinberg:2012pjx} to derive the BPS Lagrangian that would lead to these Bogomolny's equations.

Using the Bogomony's trick~\cite{Bogomolny:1975de}, one can obtain the well-known Bogomolny's equations for monopoles and dyons by completing the square in the energy density~\cite{Manton:2004tk,Weinberg:2012pjx},
\begin{equation}\label{known BPS dyon}
 E_i= \sin\theta~D_i\Phi, \qquad B_i=\cos\theta~D_i\Phi,\qquad D_0\Phi=0,\qquad V=0,
\end{equation}
with $\theta$ is a real constant.
In addition there is one constraint equation that must be considered in order to find the solutions and that is the Gauss's law constraint~\cite{Manton:2004tk,Weinberg:2012pjx},
\begin{equation}\label{Gauss Law}
 D_iF_{0i}=ie\left[\Phi,D_0\Phi\right],
\end{equation}
which is essentially the Euler-Lagrange equations for the gauge scalar potential $A_0$. Notes that the Gauss's law constraint is trivially satisfied in the BPS limit and thus we only need to consider and solve the Bogomolny's equations \eqref{known BPS dyon} in order to find the BPS monopole and dyon solutions. Using these Bogomolny's equations, we can rewrite the Lagrangian density (\ref{SU(2) YMH}) to be
\begin{eqnarray}
 \mathcal{L}&=&\tr\left(E_i-\sin\theta D_i\Phi\right)^2-\tr\left(B_i-\cos\theta D_i\Phi\right)^2+\tr\left(D_0\Phi\right)^2-V\nonumber\\
 &&+2\sin\theta~\tr\left(E_iD_i\Phi\right)-2\cos\theta~\tr\left(B_iD_i\Phi\right)-2\sin^2\theta~\tr\left(D_i\Phi\right)^2.
\end{eqnarray}
In the BPS Lagrangian method we set $\mathcal{L}-\mathcal{L}_{BPS}=0$ in the BPS limit, which is the limit where the Bogomolny's equations (\ref{known BPS dyon}) are satisfied, and thus implies the BPS Lagrangian density
\begin{equation}\label{known BPS eqns}
 \mathcal{L}_{BPS}=2\sin\theta~\tr\left(E_iD_i\Phi\right)-2\cos\theta~\tr\left(B_iD_i\Phi\right)-2\sin^2\theta~\tr\left(D_i\Phi\right)^2.
\end{equation}
So here we find that the BPS Lagrangian density consists of terms proportional to $B_iD_i\Phi,$ $E_iD_i\Phi, \text{and} \left(D_i\Phi\right)^2$. Furthermore setting all terms in $\mathcal{L}-\mathcal{L}_{BPS}$ to be zero gives us the Bogomolny's equations (\ref{known BPS dyon}) in which here their solutions shall be called the standard BPS monopoles and dyons, respectively for $\sin(\theta)=0$ and  $\sin(\theta)\neq0$.

Now let us write a slightly more general BPS Lagrangian density, than the previous one, as follows
\begin{equation}
 \mathcal{L}_{BPS}=-2\beta~\tr\left(B_iD_i\Phi\right)+2\alpha~\tr\left(E_iD_i\Phi\right)-\left(\alpha^2-\beta^2+1\right)\tr\left(D_i\Phi\right)^2,
\end{equation}
where now $\alpha \text{ and } \beta$ are arbitrary constants. We would like to prove that the Bogomolny's equations (\ref{known BPS dyon}) and also the Gaus's law constraint (\ref{Gauss Law}) can be rederived using this BPS Lagrangian density. 
Taking $\mathcal{L}-\mathcal{L}_{BPS}=0$ and setting all terms to be zero gives us Bogomolny's equations
\begin{equation}\label{BPS Eqs}
 E_i=\alpha D_i\Phi, \qquad B_i=\beta D_i\Phi,\qquad D_0\Phi=0,\qquad V=0.
\end{equation}
One can show that Euler-Lagrange equation of the first term in the BPS Lagrangian density above is trivial using the Bianchi identity $D_iB_i=0$ and a relation $\left[D_i,D_j\right]\Phi=-ie\left[F_{ij},\Phi\right]$, and hence it is indeed a ``boundary'' term while the remaining terms turn out to be ``non-boundary'' terms which contribute to the Euler-Lagrange equations of the BPS Lagrangian density: for $\Phi$,
\begin{equation}\label{CE1}
 \alpha D_iF_{0i}-\left(\alpha^2-\beta^2+1\right)D_iD_i\Phi=0,
\end{equation}
for $A_i$,
\begin{equation}\label{CE2}
 \alpha\left(D_0D_i\Phi-ie\left[F_{0i},\Phi\right]\right)=ie \left(\alpha^2-\beta^2+1\right)\left[\Phi,D_i\Phi\right],
\end{equation}
and for $A_0$,
\begin{equation}\label{CE3}
 \alpha D_iD_i\Phi=0.
\end{equation}
The equations (\ref{CE1}), (\ref{CE2}), and (\ref{CE3}) are additional constraint equations, in addition to the Bogomolny's equations (\ref{BPS Eqs}), that must be considered in finding solutions for monopoles and dyons. With these additional constraint equations, we seem to have more equations than the number of fields to be solved. In the BPS limit, in which the BPS equations (\ref{BPS Eqs}) are satisfied, these additional constraint equations can be simplified, respectively, to
\begin{subequations}
 \begin{eqnarray}
 \left(1-\beta^2\right)D_iD_i\Phi&=&0,\\
 \left(1-\alpha^2-\beta^2\right) \left[D_i\Phi,\Phi\right]&=&0,\\
 \alpha~D_iD_i\Phi&=&0,
\end{eqnarray}
\end{subequations}
where we have used the fact that $\left[D_0,D_i\right]\Phi=-ie\left[F_{0i},\Phi\right]$. We can simplify these constraint equations using the Bianchi identity $D_iB_i=0$ which, after substituting the Bogomolny's equations \eqref{BPS Eqs}, becomes $\beta D_iD_i\Phi=0$. Requiring $\beta\neq0$, the remaining constraint equation is\footnote{We can not take $\left[D_i\Phi,\Phi\right]=0$ because it could imply $[\tau^a,\tau^b]=0$, for arbitrary $a$ and $b$, which is incorrect.} 
\begin{equation}
 \left(1-\alpha^2-\beta^2\right) \left[D_i\Phi,\Phi\right]=0.
\end{equation}
Solutions to this equation is $\alpha^2+\beta^2=1$. In this BPS limit, the Gauss's law constraint \eqref{Gauss Law} is trivial and thus in finding the BPS monopoles and dyons, there are no additional equations need to be considered beside the Bogomolny's equations \eqref{BPS Eqs}. This is actually what we expected from the BPS Lagrangian method since the Bogomolny's equations \eqref{BPS Eqs} must satisfy trivially the Euler-Lagrange equations which the Gauss's law constraint is one of.

\section{Generalized BPS Monopoles and Dyons}

Following the previous sections now we may consider a more general BPS Lagrangian density to derive Bogomolny's equations for monopoles and dyons in the generalized $SU(2)$ Yang-Mills-Higgs model (\ref{gen YMH}), which is given by
\begin{equation}\label{Lbps gen}
 \mathcal{L}_{BPS}=2\alpha~\tr\left(E_iD_i\Phi\right)-2\beta~\tr\left(B_iD_i\Phi\right)-\gamma~\tr\left(D_i\Phi\right)^2,
\end{equation}
where now $\alpha\equiv\alpha(|\Phi|), \beta\equiv\beta(|\Phi|),$ and $\gamma\equiv\gamma(|\Phi|)$ are arbitrary functions of $|\Phi|$. In this case
\begin{eqnarray}\label{L-Lbps}
 \mathcal{L-L}_{BPS}&=& w~ \tr\left(E_i-\frac{\alpha}{w} D_i\Phi\right)^2 -w~ \tr\left(B_i-\frac{\beta}{w} D_i\Phi\right)^2 \nonumber\\
 &&+G~\tr\left(D_0\Phi\right)^2-\left(-\gamma +\frac{\alpha ^2}{w}-\frac{\beta ^2}{w}+G\right)\tr\left(D_i\Phi \right)^2 -V.
\end{eqnarray}
Now in the BPS limit $\mathcal{L-L}_{BPS}=0$ which implies all terms on the right hand side of (\ref{L-Lbps}) should be zero. Since $(G,w)\neq0$, the first three terms should be identified as the Bogomolny's equations
\begin{subequations}\label{BE gen}
 \begin{eqnarray}
 E_i&=&\frac{\alpha}{w} D_i\Phi,\label{BPSE gen}\\
 B_i&=&\frac{\beta}{w} D_i\Phi,\label{BPSB gen}\\
 D_0\Phi&=&0,
\end{eqnarray}
\end{subequations}
and the last term implies $V=0$. The fourth term could be zero if we set $D_i\Phi=0$, but this will make the Bogomolny's equations (\ref{BPSE gen}) and (\ref{BPSB gen}) trivial and hence $D_i\Phi\neq0$. So for this term we should take
\begin{equation}
 \gamma = G+\frac{\alpha ^2}{w}-\frac{\beta ^2}{w}.\label{BPS add}
\end{equation}
Additionally there are also constraint equations coming from the Euler-Lagrange equations of the BPS Lagrangian density (\ref{Lbps gen}), which are:
for $\Phi$,
\begin{eqnarray}
&&4{\partial\alpha\over\partial|\Phi|}\left[\tr\left(\Phi\partial_i\Phi\right)E_i-\tr\left(E_iD_i\Phi\right)\Phi\right]+\alpha D_iE_i\nonumber\\
&-&4{\partial\beta\over\partial|\Phi|}\left[\tr\left(\Phi\partial_i\Phi\right)B_i-\tr\left(D_i\Phi B_i\right)\Phi\right]\nonumber\\
&+&2{\partial\gamma\over\partial|\Phi|}\left[\tr\left(D_i\Phi\right)^2\Phi-2\tr\left(\Phi\partial_i\Phi\right)D_i\Phi\right]-\gamma D_iD_i\Phi=0,
\end{eqnarray}
for $A_i$,
\begin{eqnarray}
 &&4{\partial\alpha\over\partial|\Phi|}\tr\left(\Phi\partial_0\Phi\right)D_i\Phi+\alpha\left(D_0D_i\Phi-ie\left[E_i,\Phi\right]\right)\nonumber\\
 &+&4{\partial\beta\over\partial|\Phi|}\epsilon_{ijk}\tr\left(\Phi\partial_j\Phi\right)D_k\Phi-ie\gamma\left[\Phi,D_i\Phi\right]=0,
\end{eqnarray}
for $A_0$,
\begin{eqnarray}
 -4{\partial\alpha\over\partial|\Phi|}\tr\left(\Phi\partial_i\Phi\right)D_i\Phi-\alpha D_iD_i\Phi=0.
\end{eqnarray}
As shown in the previous section, we write these constraint equations in the BPS limit namely by substituting the Bogomolny's equations (\ref{BPSE gen}), (\ref{BPSB gen}), $D_0\Phi=0$, and $V=0$, together with the equation (\ref{BPS add}). The constraint equations are now simplified, respectively, to
\begin{subequations}\label{gen CE}
 \begin{eqnarray}
 &-&4\left(G'-{\beta\over w}\beta'+{\beta^2\over w^2}w'\right) \tr\left(\Phi D_i\Phi\right)D_i\Phi -\left(G-{\beta^2\over w}\right)D_i D_i\Phi \nonumber\\
 &+&2 \left(G'-{\alpha^2\over w^2}w'+{\beta^2\over w^2}w'\right) \tr\left(D_i\Phi\right)^2\Phi=0,\label{gen CE1}\\
 &&4\beta'\epsilon_{ijk}~\tr\left(\Phi D_j\Phi\right)D_k\Phi-ie\left(G-{\alpha^2 \over w}-{\beta^2 \over w}\right)\left[\Phi,D_i\Phi\right]=0,\label{gen CE2}\\
 &-&4\alpha'~\tr\left(\Phi D_i\Phi\right)D_i\Phi -\alpha~ D_i D_i\Phi=0,\label{gen CE3}
\end{eqnarray}
\end{subequations}
where the apostrophe $'$ means taking derivative over $|\Phi|$. 


\subsection{The Bianchi identity}
The equations of motion for the gauge fields are not only given by the Euler-Lagrange equations \eqref{full EoM gauge}, but also by the Bianchi identity,
\begin{equation}
 \epsilon^{\sigma\rho\mu\nu}D_\rho F_{\mu\nu}=0,
\end{equation}
 which can be devided into two equations
\begin{subequations}\label{BI}
 \begin{eqnarray}
 D_i B_i&=&0,\label{BI-1}\\
 2D_0 B_i&=&\epsilon_{ijk}D_{[j}E_{k]}.\label{BI-2}
\end{eqnarray}
\end{subequations}
In the BPS limit, by substituting the Bogomolny's equations \eqref{BE gen}, the equation \eqref{BI-1} becomes
\begin{equation}\label{BI-3}
 {\beta\over w} D_iD_i\Phi=-4\left(\beta\over w\right)' \tr\left(\Phi D_i\Phi\right)D_i\Phi,
\end{equation}
while, for static cases, the equation \eqref{BI-2} becomes
\begin{equation} \label{BI-2 BPS limit}
 \left(\alpha\over w\right)' \epsilon_{ijk} \tr\left(\Phi D_{[j}\Phi\right)D_{k]}\Phi=0.
\end{equation}
Using these Bianchi identity equations, the constraint equations \eqref{gen CE} can be simplified to
\begin{subequations}\label{simp gen CE}
 \begin{eqnarray}
 &-&2\left(G'-{G\over\beta}\beta'+{G\over w}w'\right) \tr\left(\Phi D_i\Phi\right)D_i\Phi+\left(G'-{\alpha^2\over w^2}w'+{\beta^2\over w^2}w'\right) \tr\left(D_i\Phi\right)^2\Phi=0,\nonumber\\\label{simp gen CE1}\\
 &&\left(\alpha\over w\right)'\left(G-{\alpha^2 \over w}-{\beta^2 \over w}\right)\left[\Phi,D_i\Phi\right]=0,\label{simp gen CE2}\\
 &&\alpha\left({\alpha'\over\alpha}-{\beta'\over\beta}+{w'\over w}\right) \tr\left(\Phi D_i\Phi\right)D_i\Phi=0.\label{simp gen CE3}
\end{eqnarray}
\end{subequations}
Non-trivial solutions require $D_i\Phi\neq0$, and so solutions to the equation \eqref{simp gen CE3} are $\alpha=0$ or $\alpha\neq0$, or to be precise $\beta=c_\beta \alpha w$, with $c_\beta$ is a constant. From now on, we require $\beta\neq0$, along with $D_i\Phi\neq0$, for BPS monopoles and dyons throughout this article, and hence $c_\beta\neq0$ .

\subsection{BPS monopoles: $\alpha=0$}

Let us first consider the case of $\alpha=0$, or $E_i=0$, which correspond to BPS monopoles case. In this case, the constraint equations (\ref{simp gen CE2}) and (\ref{simp gen CE3}) are trivially satisfied and the remaining constraint equation \eqref{simp gen CE1} can be simplified to
\begin{equation}\label{BPS mon gen CE1}
 -2\left(G'-{G\over\beta}\beta'+{G\over w}w'\right) \tr\left(\Phi D_i\Phi\right)D_i\Phi+\left(G'+{\beta^2\over w^2}w'\right) \tr\left(D_i\Phi\right)^2\Phi=0.
\end{equation}
This constraint equation can be trivial if we set $\beta=c_\beta G w$ and $w={1\over c_\beta^2 G}+c_g$, with $c_\beta\neq0$ and $c_g$ are constants. Without losing generality, we can fix these constants by comparing to the results of $SU(2)$ Yang-Mills-Higgs model, where $G=w=1$, and thus we must set $c_g=1-{1\over c_\beta^2}$. So the Bogomolny's equations for BPS monopoles are\footnote{Here the scalar potential $V$ is not absolutely zero, but it is $SU(2)$ invariant in which the model (\ref{gen YMH}) can be spontaneously broken to $U(1)$ gauge symmetry, see~\cite{Manton:2004tk,Weinberg:2012pjx} for more detail.}
\begin{equation}\label{gen BPS monopoles}
 E_i=0,\qquad B_i=c_\beta G~D_i\Phi,\qquad D_0\Phi=0,\qquad V=0
\end{equation}
with $w={1\over c_\beta^2 G}+1-{1\over c_\beta^2}$ and $c_\beta\neq0$ is a real constant. Setting $c_\beta^2=1$, we get back the results of~\cite{Casana:2012un,Atmaja:2018cod} where $w=1/G$. In general we shall call the Bogomolny's equations \eqref{gen BPS monopoles}, with(out) additional constraint equation \eqref{BPS mon gen CE1}, as the generalized Bogomolny's equations for BPS monopoles whose solutions, with $w$ or $G$ are non-constants, shall be called generalized BPS monopoles.

\subsection{BPS dyons: $\alpha\neq0$}\label{gen BPS dyons}
As we mentioned previously here $\beta= c_\beta \alpha w$. Later on, the constraint equation \eqref{simp gen CE2} implies $\alpha=c_\alpha w$, with $c_\alpha\neq0$ is a constant, or $Gw=\alpha^2+\beta^2=\alpha^2\left(1+c_\beta^2 w^2\right)$.

\subsubsection{The case of $\alpha=c_\alpha w$}\label{independent couplings}
In this case, the constraint equation \eqref{simp gen CE1} can be simplified to
\begin{equation}\label{BPS dyon gen CE1-1}
 -2\left(G'-{G\over w}w'\right) \tr\left(\Phi D_i\Phi\right)D_i\Phi+\left(G'+c_\alpha^2 \left(c_\beta^2 w^2-1\right)w'\right) \tr\left(D_i\Phi\right)^2\Phi=0
\end{equation}
which is trivially satisfied if $G$ and $w$ is constants, and thus lead us to the standard BPS dyon solutions. Therefore the generalized BPS dyons may exist as solutions to the constraint equation \eqref{BPS dyon gen CE1-1} in addition to the Bogomolny equations
\begin{equation}\label{BE gen BPS dyons}
E_i=c_\alpha D_i\Phi,\qquad B_i=c_\alpha c_\beta w~D_i\Phi,\qquad D_0\Phi=0,\qquad V=0.
\end{equation}
Here, in general, the scalar fields-dependent couplings $G$ and $w$ are independent to each other, but their relation can be determined from the above constraint equation \eqref{BPS dyon gen CE1-1}, which depend on the choice of ansatz.

\subsubsection{The case of $G={\alpha^2\over w}\left(1+c_\beta^2 w^2\right)$}\label{gen couplings relation}
In this case, the constraint equation \eqref{simp gen CE1} can be simplified to
\begin{equation}\label{stable CE}
 \left(\alpha \left(c_\beta^2 w^2-1\right) w'+w \left(c_\beta^2 w^2+1\right) \alpha'\right)\left(\tr\left(\Phi D_i\Phi\right)D_i\Phi- \tr\left(D_i\Phi\right)^2\Phi\right)=0
\end{equation}
which is trivially satisfied if $\alpha={c_\alpha w \over 1+c_\beta^2 w^2}$, with $c_\alpha\neq0$ is a constant. So we have two possible Bogomolny's equations:

\begin{itemize}
 \item $\alpha={c_\alpha w \over 1+c_\beta^2 w^2}$
 
 In this case, the generalized BPS dyon solutions can be obtained by solving the Bogomolny's equations
\begin{equation}
 E_i={c_\alpha \over 1+c_\beta^2 w^2}~ D_i\Phi,\qquad B_i={c_\alpha c_\beta w \over 1+c_\beta^2 w^2}D_i\Phi,\qquad D_0\Phi=0,\qquad V=0,
\end{equation}
 without additional constraint equation.
 
 \item $\alpha\neq{c_\alpha w \over 1+c_\beta^2 w^2}$
 
 In this case, the generalized BPS dyon solutions can be obtained by solving the Bogomolny's equations
 \begin{equation}\label{BE gen BPS dyons-1}
 E_i={\alpha \over w}~ D_i\Phi,\qquad B_i={c_\beta \alpha}D_i\Phi,\qquad D_0\Phi=0,\qquad V=0,
\end{equation}
 with additional constraint equation
 \begin{equation}\label{CE ansatz}
  \tr\left(\Phi D_i\Phi\right)D_i\Phi= \tr\left(D_i\Phi\right)^2\Phi.
 \end{equation}
 This constraint equation may only be satisfied trivially by some particular ansatze, and may not be useful to fix some of the scalar fields-dependent couplings. Since the main objective of this article is to derive Bogomolny's equations that are valid by any ansatz, and thus we may neglect this type of constraint equations.
\end{itemize}
 
Similar to the case of the $SU(2)$ Yang-Mills-Higgs model, one can easily show that the Gauss's law constraint equations of the generalized model (\ref{gen YMH}),
\begin{equation}\label{Gauss Law gen}
 4 w' \tr\left(\Phi D_i\Phi\right) E_i+w D_i E_i= ieG\left[\Phi,D_0\Phi\right],
\end{equation}
is satisfied trivially in the BPS limit for the case of BPS monopoles, by taking $\alpha=0$, and the case of BPS dyons, by taking $\beta= c_\beta \alpha w$. Here we find a more general relation between the scalar fields-dependent couplings, as shown in the section \ref{gen couplings relation} where $Gw=\alpha^2\left(1+c_\beta^2 w^2\right)$, than the one derived in the spherically symmetric system~\cite{Atmaja:2018cod}, which is a particular case with $\alpha$ is constant.

Now we will show that there are no generalized BPS dyons for the Julia-Zee ansatz (\ref{eq:ansatz}) in all of the cases described above. Let us assume general case of $\alpha\neq0$. Using the ansatz (\ref{eq:ansatz}), the Bogomolny's equation (\ref{BPSE gen}) yields
\begin{equation}
 wj+e~\alpha f=0,\qquad\qquad w{dj\over dr}+e~\alpha {df\over dr}=0,
\end{equation}
where $\alpha$ and $w$ are functions of $f$ only, whose solutions are $j=-e{\alpha\over w}f$, with ${\alpha\over w}\neq 0$ is a constant. Without losing generality, we therefore just need to consider the case of $\alpha=c_\alpha w$ as in the section \ref{independent couplings}. On the other hand the Bogomolny's equation (\ref{BPSB gen}) implies
\begin{equation}
 {da\over dr}=e~c_\alpha c_\beta a f w,\qquad\qquad {df\over dr}={a^2-1\over e~c_\alpha c_\beta r^2 w}.
\end{equation}
Substituting those Bogomolny's equations into the constraint equation \eqref{BPS dyon gen CE1-1} implies
\begin{eqnarray}
 &&\left(a^2-1\right)^2 \left({\partial w\over\partial f} \left(c_\alpha^2 w \left(c_\beta^2 w^2-1\right)+2 G\right)-w {\partial G\over \partial f}\right)\nonumber\\
 &+&2 a^2 c_\alpha^2 c_\beta^2 e^2 f^2 w^3 \left(c_\alpha^2 \left(c_\beta^2 w^2-1\right) {\partial w\over\partial f}+{\partial G\over \partial f}\right)r^2=0,
\end{eqnarray}
where $G$ is a function of $f$ only. This equation can be solved by considering it as a polynomial equation of explicit radial coordinate $r$ and then setting all its ``coefficients'' to zero,
\begin{eqnarray}
 {\partial w\over\partial f} \left(c_\alpha^2 w \left(c_\beta^2 w^2-1\right)+2 G\right)-w {\partial G\over \partial f}=0,\nonumber\\
 c_\alpha^2 \left(c_\beta^2 w^2-1\right) {\partial w\over\partial f}+{\partial G\over \partial f}=0.
\end{eqnarray}
Solutions to these equations are given by $w$ and $G$ are constants, and so we will get back the Bogomolny's equations for standard BPS dyons, whose solutions have been studied in~\cite{Prasad:1975kr,Manton:2004tk,Weinberg:2012pjx}. Here we may conclude that the Julia-Zee ansatz (\ref{eq:ansatz}) is not suitable for finding the generalized BPS dyon solutions of the Bogomolny's equations \eqref{BE gen}. However there might be generalized BPS dyon solutions under different ansatze, such as axially symmetric ansatz~\cite{Hartmann:2000ja}, which is beyond the discussion of this article and will be discussed elsewhere.

\subsection{Stress-Energy-Momentum density tensor}
Although we do not have explicit generalized BPS dyon solutions, we may still learn some of their features from the stress-energy-momentum density tensor due to the existence of the generalized Bogomolny's equations for BPS dyons.
The stress-energy-momentum density tensor of the generalized model (\ref{gen YMH}) is defined as
\begin{equation}\label{SEM Tensor}
 T_{\mu\nu}=2 G~\tr\left(D_\mu\Phi D_\nu\Phi\right)-2 w~ \tr\left(F_{\lambda\mu}F^\lambda_{~\nu}\right)-\eta_{\mu\nu}\mathcal{L}.
\end{equation}
It is usually argued that the static soliton solutions are stable if their total energy $E=\int d^3x\sqrt{-g}~ T^0_0$, in the BPS limit, is proportional to the topological charge~\cite{Bogomolny:1975de,Manton:2004tk,Weinberg:2012pjx}. Here we would like suggest that stability of the static solution solutions can be seen from their stress density tensor. Physically, the stress density tensor is related to (internal) pressures and shear stress of the solutions. It is natural to expected that all stable solutions have vanishing (internal) pressures and shear stress. Not surprisingly, one can check that all stress density tensor components of many well-known (stable) BPS solutions are zero in the BPS limit. There are some examples where the BPS solutions, with some of their stress density tensor components are non-zero, are found to be unstable either because the total static energy is not proportional to the topological charge or the solution does not exist or unphysical~\cite{Atmaja:2018ddi,Fadhilla:2020rig}. To meet this requirement, we compute the stress density tensor components of the generalized model (\ref{gen YMH}) that, in the BPS limit, is given by
\begin{equation}
 T_{ij}=2\left(G-{\alpha^2\over w}-{\beta^2\over w}\right)\tr\left(D_i\Phi D_j\Phi\right)-\delta_{ij}\left(G-{\alpha^2\over w}-{\beta^2\over w}\right)\tr\left(D_k\Phi\right)^2.
\end{equation}
Therefore the generalized BPS monopole and dyon solutions, if they exist, are stable when $Gw=\alpha^2+\beta^2$. This additional relation between the scalar fields-dependent couplings will result in stable generalized BPS monopole and dyon solutions.

\subsubsection{stable generalized BPS monopoles}
 For the case of generalized BPS monopoles, the constraint equation \eqref{BPS mon gen CE1} is simplified to
 \begin{equation}
  \beta' \left(\tr\left(\Phi D_i\Phi\right)D_i\Phi- \tr\left(D_i\Phi\right)^2\Phi\right)=0.
 \end{equation}
 Assuming $\tr\left(\Phi D_i\Phi\right)D_i\Phi\neq \tr\left(D_i\Phi\right)^2\Phi$ implies $\beta$ is constant and the relation $Gw=1$, where $\beta^2$ has been normalized to unity. The Bogomolny's equations are then given by
\begin{equation}\label{stable gen BPS monopoles}
 E_i=0,\qquad B_i=\pm G~D_i\Phi,\qquad D_0\Phi=0,\qquad V=0.
\end{equation}

\subsubsection{stable generalized BPS dyons}
 For the case of generalized BPS dyons, the remaining constraint equation \eqref{simp gen CE1} simply becomes the constraint equation \eqref{stable CE}. Again, by assuming $\tr\left(\Phi D_i\Phi\right)D_i\Phi\neq \tr\left(D_i\Phi\right)^2\Phi$, we obtain $\alpha={c_\alpha w \over 1+c_\beta^2 w^2}$ and the (normalized) relation\footnote{We normalize the constants $c_\alpha$ and $c_\beta$ by comparing with the results of $SU(2)$ Yang-Mills-Higgs model, where $G=w=1$, and so they are related by $1+c_\beta^2=c_\alpha^2$, or to be precise we chose $c_\alpha=\csc(\theta)$ and $c_\beta=\cot(\theta)$.} 
 \begin{equation}\label{stable relation}
  G={w \over \sin^2(\theta)+\cos^2(\theta)~w^2}.
 \end{equation}
The Bogomolny's equations are then given by\footnote{By including $\sin(\theta)=0$, the Bogomolny's equations \eqref{stable gen BPS dyons}, together with the (normalized) relation of the scalar fields-dependent couplings \eqref{stable relation}, are also valid for the stable generalized BPS monopoles where $\cos(\theta)=\pm 1$.}
  \begin{equation}\label{stable gen BPS dyons}
 E_i=\sin(\theta){G \over w} D_i\Phi,\qquad B_i=\cos(\theta)G~D_i\Phi,\qquad D_0\Phi=0,\qquad V=0,
\end{equation}
with $\cos(\theta)\neq0$ and $\sin(\theta)\neq0$, where $\theta$ is a real constant.

The static energy density, in the BPS limit, is given by
\begin{equation}
 T_{00}=\left(G+{\alpha^2\over w}+{\beta^2\over w}\right)\tr\left(D_i\Phi\right)^2.
\end{equation}
The (stable) generalized BPS monopole and dyon solutions have energy density
\begin{equation}
 T_{00}=2G~\tr\left(D_i\Phi\right)^2.
\end{equation}


\subsection{Topological Charge}
The integer topological charge is defined as~\cite{Weinberg:2012pjx}
\begin{equation}\label{TC}
 N_\Phi={1\over 8\pi}\int_{|x|\to\infty} dS^i\epsilon^{ijk}\epsilon^{abc}\phi^a \partial_j\phi^b \partial_k\phi^c,
\end{equation}
where $\phi\equiv{\Phi\over\sqrt{|\Phi|}}$. Non-trivial topological charge requires $|\Phi|\to v>0$ near the spatial infinity and thus typical potential for the scalar fields is $V={\lambda\over 4}\left(|\Phi|-v\right)^2$, with $\lambda=0$ in the BPS limit. Furthermore the finite energy configuration requires near the spatial infinity $D_i\Phi$ fall faster than $|x|^{-3/2}$ and in addition we also require the functions $G$ and $w$ to be finite everywhere. Under those requirements, the topological charge (\ref{TC}) can be rewritten as
\begin{equation}
 N_\Phi= -{e\over 2\pi} \int_{|x|\to\infty} dS^i~\tr\left(\phi B_i\right).
\end{equation}
In the BPS limit, by substituting $B_i$ from the equation \eqref{stable gen BPS dyons}, it becomes
\begin{equation}
 N_\Phi= -{e\cos(\theta)\over 2\pi\sqrt{v}} \int_{|x|\to\infty} dS^i~G~\tr\left(\Phi D_i\Phi\right).
\end{equation}
Now the total static energy of BPS dyon is given by
\begin{eqnarray}
 E_{BPS}&=&2\int d^3x~ G~ \tr\left(D_i\Phi\right)^2,\nonumber\\
 &=&2  \int_{|x|\to\infty} dS^i~G~\tr\left(\Phi D_i\Phi\right)-2\int d^3x~\tr\left(\Phi\left(G D_iD_i\Phi+4G'\tr\left(\Phi D_i\Phi\right)D_i\Phi\right)\right).\nonumber\\
\end{eqnarray}
The first term in the second line is obtained using the Gauss's theorem. The second term in the second line is equal to zero by the Bianchi identity equation (\ref{BI-3}), and thus the total static energy of BPS dyon is proportional to the integer topological charge,
\begin{equation}
 E_{BPS}=4\pi\sqrt{v}\left|N_\Phi\over e\cos(\theta)\right|.
\end{equation}
Here, we show that solutions to the BPS dyon equations \eqref{stable gen BPS dyons} are indeed stable.

\section{Conclusions}
We made use of the well-known Bogomolny's equations (\ref{known BPS dyon}) in the $SU(2)$ Yang-Mills-Higgs model in order to get all possible terms of the correponding BPS Lagrangian density \eqref{known BPS eqns} that would lead to these Bogomolny's equations. We generalized this BPS Lagrangian density by multiplying each of its terms with arbitrary function of the scalar fields, and then used the generalized BPS Lagrangian density (\ref{Lbps gen}) to derive the generalized Bogomolny's equations in the generalized $SU(2)$ Yang-Mills-Higgs model (\ref{gen YMH}). In the case of BPS monopoles, the generalized Bogomolny's equations are given by the equations \eqref{gen BPS monopoles} with(out) additional constraint equation \eqref{BPS mon gen CE1}. In the case of BPS dyons, there are two possible generalized Bogomolny's equations which are given by the equations \eqref{BE gen BPS dyons} with(out) additional constraint equation \eqref{BPS dyon gen CE1-1}, and by the equations \eqref{BE gen BPS dyons-1} with(out) additional constraint equation \eqref{stable CE}. We then argued that the BPS monopole and dyon solutions to those generalized Bogomolny's equations are stable if all components of the stress density tensor are zero. This additional stability requirement yields the generalized Bogomolny's equations \eqref{stable gen BPS dyons} and the equation \eqref{stable relation} relating the scalar fields-dependent couplings for BPS monopoles, where $\sin(\theta)=0$, and for BPS dyons without any constraint equation. Moreover we showed that the total energy of BPS dyon is proportional to the topological charge, $E_{BPS}\propto |N_\Phi|$.

It is easy to show that by substituting the 't Hooft-Polyakov ansatz, which is (\ref{eq:ansatz}) with $j=0$, into the generalized Bogomolny's equations \eqref{stable gen BPS monopoles} we will get back the self-dual (BPS) equations, and also the same equation relating the scalar fields-dependent couplings, whose solutions are the generalized BPS monopoles studied in~\cite{Casana:2012un}. Therefore the generalized Bogomolny's equations \eqref{stable gen BPS monopoles} are indeed the general coordinate extension of those self-dual (BPS) equations. Unfortunately, for the case of BPS dyons, substituting the Julia-Zee ansatz \eqref{eq:ansatz} into the generalized Bogomolny's equations \eqref{stable gen BPS dyons} implies $G\propto w$ which then, from the equation \eqref{stable relation}, further imply $\cos(\theta)=0$, or $B_i=0$. There are possible BPS dyon solutions, with $\cos(\theta)\neq0$, if both $w$ and $G$ are constants, or known as the standard BPS dyons, and therefore there are no generalized Bogomolny's equations for BPS dyons under the Julia-Zee ansatz. However this contradicts with the results in~\cite{Atmaja:2018cod} where there exist (spherically symmetric) generalized Bogomolny's equations for BPS dyons under the Julia-Zee ansatz. We also found the relation equation \eqref{stable relation} is different from the one obtained in~\cite{Atmaja:2018cod}. These contradictions appear because the computations did in~\cite{Atmaja:2018cod} is in the effective description under the Julia-Zee ansatz and the potential scalar $A^a_0$ is directly identified with the scalar fields $\Phi^a$, or namely by taking $j\propto f$, in the effective Lagrangian density \eqref{gen eff L}. In this  way, the corresponding effective Gauss's law constraint equation of the equation \eqref{Gauss Law gen} disappears from the Euler-Lagrange equations of the effective Lagrangian density \eqref{gen eff L}. Furthermore, there will be no correponding effective constraint equation of the equation \eqref{simp gen CE3}. Therefore the computations, for BPS dyons, in~\cite{Atmaja:2018cod} are actually incomplete. In Appendix \ref{Appendix}, we repeated the (effective description) computations in~\cite{Atmaja:2018cod}, for BPS dyons in the generalized model \eqref{gen YMH} under the Julia-Zee ansatz \eqref{eq:ansatz}, without first taking the identification $j\propto f$. We arrived at the same conclusion that there are no generalized Bogomolny's equations for BPS dyons and thus no generalized BPS dyon solutions.

We could consider more general BPS Lagrangian density than \eqref{Lbps gen} that could lead to the generalized BPS dyons solutions under the Julia-Zee ansatz. The BPS Lagrangian density (\ref{Lbps gen}) is not the most general BPS Lagrangian density. There are other possible terms, in terms of $E_i,B_i,D_i\Phi,$ and $D_0\Phi$, that one can add to the BPS Lagrangian density (\ref{Lbps gen}). The first one is a term that are independent to all first-derivative of the fields, or basically an arbitrary function of the scalar fields $|\Phi|$. The second one, a term that is proportional to $\tr\left(D_0\Phi\right)$. The third ones are the remaining terms that are proportional to quadratic of first-derivative of the fields: $\tr\left(E_i\right)^2, \tr\left(B_i\right)^2, \tr\left(D_0\Phi\right)^2,$ and $\tr\left(E_iB_i\right)$. Another possible way to find the generalized BPS dyons is by considering different ansatze such as axially symmetric ansatz studied in~\cite{Hartmann:2000ja} for higher topological charge dyons. However those possibilities are beyond the study of this article and they will be investigated elsewheres.

\acknowledgments
I would like to acknowledge the Abdus Salam ICTP for Associateships 2019 and for warmest hospitality during the Associate visit where this work was initially done. 
This work was supported under the grant WCR Kemenristek 2022.

\appendix

\section{Generalized BPS dyons in the effective description: spherically symmetric}\label{Appendix}
In the effective description, we will apply the Julia-Zee ansatz (\ref{eq:ansatz}) directly into the Lagrangian density of generalized $SU(2)$ Yang-Mills-Higgs model (\ref{gen YMH}) such that the effective Lagrangian density is spherically symmetric and is given by
\begin{equation}\label{gen eff L}
 \mathcal{L}_{eff}=-G(f) \left(\frac{f'^2}{2}+\frac{a^2 f^2}{r^2}\right)+{w(f)\over e^2} \left(\frac{j'^2}{2}+\frac{a^2 j^2}{r^2}\right)-{w(f)\over e^2} \left(\frac{a'^2}{r^2}+\frac{\left(a^2-1\right)^2}{2 r^4}\right)-V(f),
\end{equation}
where form now on the apostrophe $'$ means taking derivative over radial coordinate $r$, or over its argument if explicitly written. Here we will consider two forms of effective BPS Lagrangian density. The first one is suggested by the form of generalized BPS Lagrangian density \eqref{Lbps gen} and the second one is following the form in~\cite{Atmaja:2018cod} which is linear to all first-derivative of the effective fields.

\subsection{Effective BPS Lagrangian density motivated by the BPS Lagrangian density \eqref{Lbps gen}}\label{A1}
Under the Julia-Zee anstaz (\ref{eq:ansatz}), each terms in the generalized BPS Lagrangian density \eqref{Lbps gen} are proportional to the first-derivative of the effective fields as follow
\begin{subequations}
 \begin{align}
  \tr\left(E_i D_i\Phi\right)&\propto f'j'\\
  \tr\left(B_i D_i\Phi\right)&\propto {a'\over r^2},\mbox{ and}~{f'\over r^2}\\
  \tr\left(D_i\Phi\right)^2&\propto f'^2.
 \end{align}
\end{subequations}
This then suggests that we should take the following effective BPS Lagrangian density\footnote{Since the effective Lagrangian density (\ref{gen eff L}) is spherically symmetric, the effective BPS Lagrangian density should also be spherically symmetric as well.}
\begin{equation}
 \mathcal{L}_{BPS}=-Q_f(f,a,j) {f'\over r^2}-Q_a(f,a,j) {a'\over r^2}-X_1(f,a,j) f'j'-X_2(f,a,j) f'^2,\label{L_BPS eff}
\end{equation}
where $Q_f,Q_a,X_1,X_2$ are arbitrary functions of the effective fields with no explicit radial coordinate dependent.
Solving $\mathcal{L}_{eff}-\mathcal{L}_{BPS}=0$ implies Bogomolny's equations: for $f$,
\begin{subequations}\label{BE eff}
\begin{equation}
  f'={Q_f+r^2 X_1 j' \over r^2 \left(G-2X_2\right)};
\end{equation}
for $a$, \begin{equation}
          a'={e^2\over 2} {Q_a\over w};
         \end{equation}
and for $j$, \begin{equation}
              j'=-e^2 {Q_f X_1 \over r^2\left(w\left(G-2X_2\right)+e^2 X_1^2\right)}.
             \end{equation}
\end{subequations}
The residual equation of $\mathcal{L}_{eff}-\mathcal{L}_{BPS}=0$ is
\begin{equation}
\frac{w\left(e^2 Q_f^2-\left(a^2-1\right)^2 \left(e^2 X_1^2+w (G-2 X_2)\right)\right)} {2r^4e^2 \left(e^2 X_1^2+w (G-2 X_2)\right)}+\frac{4 a^2 w \left(j^2 w-e^2 f^2 G\right)+e^4 Q_a^2}{4r^2 e^2 w}=V
\end{equation}
which can be solved by setting all the ``coefficients'' in its explicit radial coordinate $r$ expansion to be zero such that $V=0$,
\begin{equation}
 Q_a=\pm {2\over e^2} a \sqrt{w \left(e^2f^2 G-j^2 w\right)},
\end{equation}
and
\begin{equation}
 X_2=\frac{e^2}{2} \left(\frac{ \left(a^2-1\right)^2 X_1^2-Q_f^2}{\left(a^2-1\right)^2 w}+{G\over e^2}\right).
\end{equation}

Now, there are still two remaining functions ($Q_f$ and $X_1$) need to be determined by using Euler-Lagrange equations of the BPS Lagrangian density (\ref{L_BPS eff}) for $f,a,$ and $j$;
\begin{equation}
 \partial_r \left(\partial (r^2\mathcal{L}_{BPS}) \over \partial_r\phi\right)-{\partial (r^2\mathcal{L}_{BPS}) \over \partial\phi}=0,
\end{equation}
where $\phi\equiv(f,a,j)$ is the effective field.
Substituting all Bogomolny's equations (\ref{BE eff}) and explicit solutions for $V,Q_a,$ and $X_2$ into those Euler-Lagrange equations, that shall now be called as constraint equations, then we can solve them similarly by setting all the ``coefficients'' in their explicit radial coordinate $r$ expansion to be zero. In this way, each of the constraint equations can be written into terms with explicit power of radial coordinate $r$: $r^{-2}$- and $r^0$-terms. 
The function $Q_f$ can determined from the $r^0$-term of the constraint equation for $f$ by equation
\begin{eqnarray}
&& a^2 e^2 f {G\over w} \left(f w'(f)+2w\right)\pm\frac{a G}{Q^2_f}\left(\left(a^2-1\right) {\partial Q_f\over\partial a}- 4aQ_f\right)\left(a^2-1\right) \sqrt{w \left(e^2f^2 G-j^2 w\right)}\nonumber\\
 &+&a^2 \left(e^2f^2 G'(f)-2 j^2 w'(f)\right)=0,
\end{eqnarray}
which has solution
\begin{equation}
 Q_f=\pm\frac{2\left(a^2-1\right) G w \sqrt{w \left(e^2f^2 G-j^2 w\right)}}{w \left(e^2f^2 G'(f)-2 j^2 w'(f)\right)+e^2f G \left(f w'(f)+2 w\right)}.
\end{equation}
Using these solutions for $Q_f$, the function $X_1$ can be determined from the $r^{-2}$-term of the constraint equation for $a$,
\begin{eqnarray}
 X_1&=&\frac{1}{2 j}\left(\frac{4 G w \left(e^2f^2 G-j^2 w\right)}{w'(f) \left(e^2f^2 G-2 j^2 w\right)+e^2f w \left(f G'(f)+2 G\right)}+\frac{w'(f) \left(2 j^2 w-e^2f^2 G\right)}{e^2w}\right. \nonumber\\
 &&\qquad\left.-f \left(f G'(f)+2 G\right)\right),
\end{eqnarray}
and also from the $r^{-2}$-term of the constraint equation for $j$,
\begin{equation}
 X_1=-\frac{2 j G w^2}{w'(f) \left(e^2f^2 G-2 j^2 w\right)+e^2f w \left(f G'+2 G\right)}.
\end{equation}
Both solutions are equal if $w=cw^2$, and $G=cg^2$ or $G={cg^2\over f^4}$, where $cw^2$ and $cg^2$ are non-zero real constants. Substituting them then all the remaining constraint equations are trivially satisfied. So we have two possible solutions, with $w=cw^2$,
\begin{enumerate}
 \item $G=cg^2$\\
  Bogomolny's equations are given by
  \begin{subequations}\label{standard BE}
  \begin{eqnarray}
   f'(r)&=&\pm\frac{\left(a(r)^2-1\right) f(r)}{r^2}{cw^2 \over \sqrt{cw^2 cg^2 e^2 f(r)^2- cw^4 j(r)^2}},\label{standard BE f}\\
   a'(r)&=& \pm {a(r)\over cw^2} \sqrt{cw^2 cg^2 e^2 f(r)^2- cw^4 j(r)^2},\\
   j'(r)&=&\pm\frac{\left(a(r)^2-1\right) j(r)}{r^2}{cw^2 \over \sqrt{cw^2 cg^2 e^2 f(r)^2- cw^4 j(r)^2}}.\label{standard BE j}
  \end{eqnarray}
 \end{subequations}
 Comparing the equations (\ref{standard BE f}) with (\ref{standard BE j}), we may take $j=\sigma f$, where $\sigma$ is non-zero real constant.
 
 \item $G={cg^2/ f^4}$\\
 Bogomolny's equations are given by
  \begin{subequations}\label{gen BE}
  \begin{eqnarray}
   f'(r)&=&\mp\frac{\left(a(r)^2-1\right) f(r)}{r^2}{cw^2 \over \sqrt{{cw^2 cg^2 e^2\over f(r)^2}- cw^4 j(r)^2}},\label{gen BE f}\\
   a'(r)&=& \pm {a(r)\over cw^2} \sqrt{{cw^2 cg^2 e^2\over f(r)^2}- cw^4 j(r)^2},\\
   j'(r)&=&\pm\frac{\left(a(r)^2-1\right) j(r)}{r^2}{cw^2 \over \sqrt{{cw^2 cg^2 e^2\over f(r)^2}- cw^4 j(r)^2}}.\label{gen BE j}
  \end{eqnarray}
 \end{subequations}
 Comparing the equations (\ref{gen BE f}) with (\ref{gen BE j}), we may take $j=\sigma/ f$, where $\sigma$ is also a non-zero real constant.
\end{enumerate}
One can easily show that both Bogomolny's equations, \eqref{standard BE} and \eqref{gen BE}, trivially satisfy the full Euler-Lagrange equations \eqref{full EoM} under the Julian-Zee ansatz \eqref{eq:ansatz}.

\subsection{Linear effective BPS Lagrangian density}
It is perhaps more suggestive to have effective Lagrangian density that is linear to the first-derivative of the effective fields, $\int d^3x\sqrt{-g}\mathcal{L}_{BPS}=\int dQ(f,a,j)$, as such its Euler-Lagrange equations are trivial and thus no additional constraint equations needed. In the earlier development of the BPS Lagrangian method, some of the well-known Bogomolny's equations, for BPS vortices, were rederived from this linear effective BPS Lagrangian density~\cite{Atmaja:2015umo}. Later on, it was shown that the Bogomolny's equations for BPS monopoles and dyons in the $SU(2)$ Yang-Mills-Higgs model also can be rederived from the linear effective BPS Lagrangian as well~\cite{Atmaja:2018cod}.
In this case the effective BPS Lagrangian density takes the following form:
\begin{equation}
 \mathcal{L}_{BPS}=- {Q_f(f,a,j)\over r^2} f'- {Q_a(f,a,j)\over r^2} a'- {Q_j(f,a,j)\over r^2} j',\label{L_BPS eff-1}
\end{equation}
where $Q_f, Q_a, Q_j$ are arbitrary functions. In order for the Euler-Lagrange equations of the effective BPS Lagrangian density to be trivial, the arbitrary functions must be related to each other as such $Q_f\equiv {\partial Q\over\partial f}, Q_a\equiv {\partial Q\over\partial a},$ and $Q_j\equiv {\partial Q\over\partial j}$, with $Q(f,a,j)$. However we first set them to be independent and later we will find the function $Q$.
Solving $\mathcal{L}_{eff}-\mathcal{L}_{BPS}=0$ implies Bogomolny's equations: for $f$,
\begin{subequations}\label{BE eff-1}
\begin{equation}
  f'={Q_f\over r^2 G};
\end{equation}
for $a$, \begin{equation}
          a'={e^2\over 2} {Q_a\over w};
         \end{equation}
and for $j$, \begin{equation}
              j'=-e^2 {Q_j \over r^2 w}.
             \end{equation}
\end{subequations}
The residual equation of $\mathcal{L}_{eff}-\mathcal{L}_{BPS}=0$ is 
\begin{equation}
\frac{\left(e^2 Q_f^2 w-e^4 Q_j^2 G-\left(a^2-1\right)^2 w^2 G\right)} {2r^4e^2 G w}+\frac{4 a^2 w  \left(j^2 w-e^2 f^2 G\right)+e^4 Q_a^2}{4r^2 e^2 w }=V
\end{equation}
which can be solved by setting all the ``coefficients'' in its explicit radial coordinate $r$ expansion to be zero such that $V=0$,
\begin{equation}
 Q_a=\pm {2\over e^2} a \sqrt{w \left(e^2f^2 G-j^2 w\right)},
\end{equation}
and
\begin{equation}\label{rem cond}
 e^2 Q_f^2 w-e^4 Q_j^2 G=\left(a^2-1\right)^2 w^2 G.
\end{equation}

To fix the functions $Q_f$ and $Qj$, we solve the Euler-Lagrange equations of the BPS Lagrangian density (\ref{L_BPS eff-1}) for $f,a,$ and $j$. Substituting all Bogomolny's equations (\ref{BE eff-1}) and explicit solutions of $Q_a$ into those Euler-Lagrange equations, that shall be called constraint equations, then we can solved them similarly by setting all the ``coefficients'' in their explicit radial coordinate $r$ expansion to be zero. In this way, each of the constraint equations can be divided into terms with explicit power of radial coordinate $r$: $r^{-4}$-, $r^{-2}$-, and $r^0$-terms. From the $r^0$-term of the constraint equation for $f$, we have
\begin{equation}
 e^2 {\partial Q_f\over\partial a} \sqrt{w \left(e^2 f^2 G-j^2 w\right)}=\pm a w \left(e^2 f^2 G'(f)-2 j^2 w'(f)\right)\pm a e^2 f G \left(f w'(f)+2 w\right)
\end{equation}
that gives general solutions to $Q_f$,
\begin{equation}
 Q_f=\pm\left(a^2-c_f(f,j)\right)\frac{\left(w'(f) \left(e^2 f^2 G-2 j^2 w\right)+e^2 f w \left(f G'(f)+2 G\right)\right)}{2 e^2 \sqrt{w \left(e^2 f^2 G-j^2 w\right)}},
\end{equation}
where $c_f$ is an arbitrary function of $f$ and $j$. Meanwhile, from the $r^0$-term of the constraint equation for $j$, we have
\begin{equation}
 2 a^2 j w^2\pm e^2 a  \sqrt{w \left(e^2 f^2 G-j^2 w\right)}{\partial Q_j\over\partial a}=0,
\end{equation}
that gives general solutions to $Q_j$,
\begin{equation}
 Q_j=\mp\left(a^2-c_j(f,j)\right)\frac{j w^2}{e^2 \sqrt{w \left(e^2 f^2 G-j^2 w\right)}},
\end{equation}
where $c_f$ is an arbitrary function of $f$ and $j$. Now we have explicit $a$-dependent in all of the functions $Q_f, Q_a,$ and $Q_j$. Therefore we can further expand all the constraint equations in the explicit $a$ expansion and then solve them by setting all ``coefficients'' of explicit $r$ coordinate and $a$ expansions to zero. From the $a^4$-term of the equation \eqref{rem cond}, we have
\begin{equation}
 \left(w'(f) \left(e^2 f^2 G-2 j^2 w\right)+e^2 f w \left(f G'(f)+2 G\right)\right)^2-4 e^2 j^2 G  w^3=4e^2 G w^2\left(e^2 f^2 G-2 j^2 w\right).
\end{equation}
Assuming $f$ and $j$ are independent each other, we solve it by rewriting it in the explicit $j$ expansion and then setting all its``coefficients'' to zero. Its $j^4-$ term implies $w= c_w^2$, where $c_w\neq0$ is a real constant. Fixing the function $w$, its $j^0$-term implies $G=c_g^2$ or $G=c_g^2/f^4$, where $c_g\neq0$ is a real constant. Substituting these solutions for $w$ and $G$ into the $a^2$-term of the equation \eqref{rem cond}, we have two possible solutions for $c_f$: if $G=c_g^2$ then
\begin{equation}
 c_f=\frac{c_w^2 j^2 c_j(f,j)+c_g^2 e^2 f^2-c_w^2 j^2}{c_g^2 e^2 f^2},
\end{equation}
and if $G=c_g^2/f^4$ then
\begin{equation}
 c_f=\frac{c_w^2 f^2 j^2 c_j(f,j)+c_g^2 e^2-c_w^2 f^2 j^2}{c_g^2 e^2}.
\end{equation}
It turns out from the $a^0$-term of the equation \eqref{rem cond}, with $w=c_w^2$,
\begin{equation}
 \frac{e^2 f^2 \left(f G'(f)+2 G(f)\right)^2 c_f(f,j)^2-4 c_w^2 j^2 G(f) c_j(f,j)^2}{G(f) \left(e^2 f^2 G(f)-c_w^2 j^2\right)}-4=0,
\end{equation}
both solutions of $G$, and also $c_f$, above yield the same $c_j=1$ which surprisingly also leads to the same $c_f=1$ for both. Now, we can compute the function $Q$ which is given by $Q=\pm {1\over e^2}(a^2-1)\sqrt{w \left(e^2f^2 G-j^2 w\right)}$. At the end we will get the same Bogomolny's equations \eqref{standard BE} and \eqref{gen BE} respectively for $G=c_g^2$ and $G=c_g^2/f^4$. As discussed previously at the end of section \ref{A1}, only the Bogomolny's equations \eqref{standard BE}, with $cw^2=cg^2=1$, satisfy the full Euler-Lagrange equations.

\subsection{Solutions}

\begin{enumerate}
 \item $G=cg^2$ \\
 Solutions to the Bogomolny's equations \eqref{standard BE} are
 \begin{subequations}
  \begin{align}
   f(r)&=\pm \left({\omega^2\over r}-\coth\left(r\over\omega^2\right)\right),\\
   a(r)&= {r\over \omega^2} {1\over\sinh\left(r\over\omega^2\right)},\\
   j(r)&=\sigma f(r),
  \end{align}
   \end{subequations}
  with $\omega^2={cw^2\over\sqrt{cw^2cg^2 e^2-cw^4\sigma^2}}$ and $\sigma<\left|{cg\over cw}e\right|$ is a real constant. Normalized $cw^2=cg^2=1$ we get back the standard BPS dyon solutions~\cite{Prasad:1975kr,Manton:2004tk,Weinberg:2012pjx}.

  \item $G=cg^2/f^4$\\
 Solutions to the Bogomolny's equations \eqref{gen BE} are
 \begin{subequations}
  \begin{align}
   f(r)&=\pm \left({\omega^2\over r}-\coth\left(r\over\omega^2\right)\right)^{-1},\\
   a(r)&= {r\over \omega^2} {1\over\sinh\left(r\over\omega^2\right)},\\
   j(r)&={\sigma\over f(r)},
  \end{align}
 \end{subequations}
  with $\omega^2={cw^2\over\sqrt{cw^2cg^2 e^2-cw^4\sigma^2}}$ and $\sigma<\left|{cg\over cw}e\right|$ is a real constant. These solutions imply only static energy density component of the stress-energy-momentum density tensor \eqref{SEM Tensor} is non-zero and is given by $E_{static}= 4\pi\omega^2$. Although the total static energy is finite, unfortunately the solutions for $f(r)$ are singular near the origin as such $f(r\to 0)\to\mp\infty$, and so they are unphysical\footnote{Under the Julia-Zee ansatz \eqref{eq:ansatz}, we expect that at the origin, $r=0$, scalar fields $\Phi^a$ to be single-valued as such $f(0)=0$.}. Furthermore, the Bogomolny's equations \eqref{gen BE} are not equivalent with the Bogomolny's equations \eqref{BE gen}, under the Julia-Zee ansatz, since $j\propto 1/f$. This also means that the general coordinate extension of the Bogomolny's equations \eqref{gen BE} may come from a different BPS Lagrangian density than the generalized BPS Lagrangian density \eqref{Lbps gen}. Notice that the effective BPS Lagrangian density \eqref{L_BPS eff} is slightly more general than the generalized BPS Lagrangian density \eqref{Lbps gen} which under the Julia-Zee ansatz is given by
\begin{equation}
 \mathcal{L}_{BPS}\approx-2{\beta (f)\over e}\left(a^2-1\right)\frac{f'}{r^2}-4 {\beta (f)\over e}a f \frac{a'}{r^2}-2\frac{\alpha (f)}{e}f'j'-\gamma(f) f'^2.
\end{equation}
  Considering the most general BPS Lagrangian density would result in most general Bogomolny's equations, than the Bogomolny's equations \eqref{BE gen}, that may have physical generalized BPS dyon solutions under the Julia-Zee ansatz.
\end{enumerate}

\medskip
\bibliographystyle{JHEP}
\bibliography{dyons}

\end{document}